\documentclass[a4paper]{article}
\usepackage{INTERSPEECH2022}
\usepackage{multirow}

\usepackage{color,soul}

\title{Self-supervised learning for robust voice cloning}
\name{Konstantinos Klapsas$^1$, Nikolaos Ellinas$^1$, Karolos Nikitaras$^1$, Georgios Vamvoukakis$^1$, Panos Kakoulidis$^1$, Konstantinos Markopoulos$^1$, Spyros Raptis$^1$, June Sig Sung$^2$, Gunu Jho$^2$, Aimilios Chalamandaris$^1$, Pirros Tsiakoulis$^1$}
\address{
  $^1$Innoetics, Samsung Electronics, Greece\\
  $^2$Mobile Communications Business, Samsung Electronics, Republic of Korea}
\email{\{n.ellinas, g.vamvouk, p.kakoulidis, k.markop, s.raptis, js6.sung, gunu.jho,  aimilios.ch, p.tsiakoulis\} @samsung.com, \\
\{k.klapsas, k.nikitaras\} @partner.samsung.com}

\begin{document}

\maketitle
\begin{abstract}
Voice cloning is a difficult task which requires robust and informative features incorporated in a high quality TTS system in order to effectively copy an unseen speaker’s voice.
In our work, we utilize features learned in a self-supervised framework via the Bootstrap Your Own Latent (BYOL) method, which is shown to produce high quality speech representations when specific audio augmentations are applied to the vanilla algorithm.
We further extend the augmentations in the training procedure to aid the resulting features to capture the speaker identity and to make them robust to noise and acoustic conditions.
The learned features are used as pre-trained utterance-level embeddings and as inputs to a Non-Attentive Tacotron based architecture, aiming to achieve multispeaker speech synthesis without utilizing additional speaker features.
This method enables us to train our model in an unlabeled multispeaker dataset as well as use unseen speaker embeddings to copy a speaker’s voice.
Subjective and objective evaluations are used to validate the proposed model, as well as the robustness to the acoustic conditions of the target utterance.
\end{abstract}

\section{Introduction}

Speech synthesis research has advanced from high quality attentive acoustic models, like Tacotron \cite{tacotron,tacotron2}, to state-of-the-art models that do not depend on attention, like Non-attentive Tacotron \cite{non_attentive} and FastSpeech \cite{ren2019fastspeech}.
By providing the acoustic model with learnable speaker embeddings it is shown that multispeaker speech synthesis is feasible with high fidelity \cite{ping2017deep,chen2020multispeech} and can be extended with few modifications to multilingual datasets \cite{zhang2019learning}.

The task of cloning the voice of speakers not included in the training set, requires representations that can generalize to unseen data.
Especially interesting are cases where the unseen speaker's utterances are limited \cite{arik2018neural}.
Other parameters such as speaker style cloning can also be factored in this task \cite{xie2021multi}.

\subsection{Related Work}

Since neural Text-to-Speech (TTS) models are capable of producing speech conditional to several factors, a speaker embedding is an adequate representation in order to differentiate between various speakers in a single model.
There are many ways of adapting a multispeaker model to a new speaker, for example fine-tuning \cite{taigman2017voiceloop,chen2018sample} is a standard approach that uses the target speaker's data to continue training of the base model.
In \cite{luong2020nautilus} a multi-stage speaker adaptation method is also proposed, whereas in \cite{huang2021meta} meta-learning is used in order to increase the generalization capability of the model.
Adaptation is also shown to work effectively in multilingual setups \cite{maniati21_interspeech,casanova2021yourtts}.

Alternatively, a speaker encoder that can directly predict a speaker embedding from audio can be applied with success to limited data scenarios \cite{arik2018neural}.
There is a lot of research on generalized speaker representations applicable to many other tasks, such as x-vectors for speaker recognition \cite{snyder2018x} and d-vectors for speaker verification \cite{Wan2018GeneralizedEL}.
These can be successfully applied in the cloning task \cite{d-vectors2} by using them as a pre-trained model in a transfer learning approach.
Instead of a fixed speaker embedding, in \cite{Attentron} a variable-length embedding is extracted from audio on-the-fly and conditions the decoder on speaker-dependent features producing high quality results.
The speaker identity can also be sensitive to acoustic conditions \cite{noise_taco}, so domain adversarial training can be employed in speaker adaptation and speaker encoding \cite{cong2020data} in order to perform voice cloning from noisy samples.

Self-supervised learning is a machine learning approach where the model trains itself by leveraging part of the data to generate supervisory signals for the task at hand. Pre-training methods for speech are used to obtain representations that are useful for a variety of possible tasks \cite{wang2021towards}.
State-of-the-art models like wav2vec 2.0 \cite{baevski2020wav2vec} follow a contrastive approach, while we focus on the non-contrastive methods which do not include negative samples. 
Specifically, BYOL \cite{byol} and SimSiam \cite{siamese} pair the representations learned by two neural networks by applying augmentations on the training data.
BYOL has been adapted for audio data \cite{Niizumi2021BYOLFA} in order to generate general purpose embeddings, which are shown to be useful in a large variety of tasks. We aim at applying and extending this method to the task of voice cloning and we show that it allows learning of meaningful speaker representations that can directly condition a high quality TTS model.


\subsection{Our Contributions}
To the best of our knowledge, this is the first application of learned self-supervised features on neural voice cloning.
The main contributions of the paper are as follows:
\begin{itemize}
    \item We present a voice cloning algorithm that can be trained on an unlabeled dataset with an arbitrary number of speakers.
    \item We demonstrate that our model is able to perform voice cloning with similar performance to a d-vectors baseline while using only a fraction of the training dataset. 
    \item We incorporate additional augmentations in order to make the learned self-supervised representations robust to acoustic conditions and prosodic variations, enabling robust voice cloning.
\end{itemize}
\section{Method}

\subsection{Overview}
Our approach is based on a Non-attentive Tacotron TTS architecture \cite{non_attentive}, adapted for producing features for the LPCNet vocoder \cite{lpcnet} and conditioned on pre-trained self-supervised features.
The pre-training method we use in our work is an adaptation of the Bootstrap Your Own Latent (BYOL) algorithm \cite{byol}, an effective self-supervised learning method that produces meaningful representations. 
In previous work, the original algorithm was adapted for learning audio representations by introducing audio-related augmentations and was called BYOL for Audio (BYOL-A) \cite{Niizumi2021BYOLFA}.
Given that this algorithm has been shown to be effective in the task of speaker identification without a labeled dataset, we simply condition our TTS system on the audio representations as speaker embeddings to achieve voice cloning.


In the following paragraphs, the training algorithm is explained in brief, along with the additional augmentations we included to better help the model capture speaker identity, and make it more robust to noise in the reference samples.

\subsection{BYOL for audio}
BYOL training consists of the simultaneous training of two neural networks, the online and target networks. 
The two networks have the same architecture but use a different set of weights, denoted as $\theta$ for the online and as $\xi$ for the target network.
Specifically, both networks consist of a representation encoder $f$, and a projector $g$, namely $f_\theta, g_\theta$  for the online and $f_\xi, g_\xi$ for the target network. Additionally, the online network has an additional prediction module $q_\xi$.

The training is done by first producing two augmented views of the audio $x$, $u \triangleq t(x)$ and $u^{\prime} \triangleq t^{\prime}(x)$ where $t$ and $t^{\prime}$ are sampled from the distribution of audio augmentations $t, t^{\prime} \sim T$.
The online network is then used to output the representation $y_{\theta} \triangleq f_{\theta}(u)$, and a projection $z_{\theta} \triangleq g_{\theta}(y)$.
It is important to note that during inference, only the representation $y_{\theta} = f_{\theta}(u)$ is used.

Similarly, the target network is used to output the target projection $z^{\prime}_{\xi} \triangleq g_{\xi}(f_{\xi}(u^{\prime}))$ from the second augmented view. The additional prediction module is used on the online prediction to get the prediction $q_{\theta}(z_{\theta})$ of $z^{\prime}_{\xi}$, which is $\ell_2$ normalized along with the target projection, $\bar{q}_{\theta}(z_{\theta}) \triangleq q_{\theta}(z_{\theta}) / \| q_{\theta}(z_{\theta}) \|_2 $ and $\bar{z}^{\prime}_{\xi} \triangleq z^{\prime}_{\xi} / \| z^{\prime}_{\xi} \|_2 $.

The loss is defined as:
\begin{equation}\label{eq:byol_loss}
    \mathcal{L}_{\theta, \xi} \triangleq \|\bar{q}_{\theta}(z_{\theta}) - \bar{z}^{\prime}_{\xi} \|_2
\end{equation}
In order for the loss to be symmetric with respect to the augmentations, the $u^{\prime}$ augmentation is also fed to the  online network and $u$ to the target network and the loss is recomputed to get $\mathcal{L}^{\sim}_{\theta, \xi}$. 
The final loss is $\mathcal{L}^{BYOL}_{\theta, \xi} = \mathcal{L}^{\sim}_{\theta, \xi} + \mathcal{L}_{\theta, \xi}$.

Only the online network is updated in order to minimize this training loss. 
The parameters of the target network are updated as an exponential moving average of the online parameters \cite{update_target}, as follows:

\begin{equation}\label{eq:target}
    \xi \leftarrow	\tau \xi + ( 1 - \tau ) \theta 
\end{equation}
where $\tau \in [0,1]$ is the target decay rate, which is set to $0.99$ in our experiments. 
It has been shown \cite{byol, Tian2021UnderstandingSL} that this training procedure is sufficient to avoid collapsed solutions such as constant representations, since the updates to the target network parameters $\xi$ are not in general in the direction of $ \nabla_{\xi} \mathcal{L}^{BYOL}_{\theta, \xi}$, due to the stop grad operation in the target network.

The input to the networks in the case of audio data is a one second segment of log-mel spectrograms. 


\subsection{BYOL-A Augmentations}
\subsubsection{Pre and post Normalization}
Both pre and post-augmentation normalization is done on the samples, $\Tilde{x} = \frac{x - \mu}{\sigma}$ where $\mu$ is the mean and $\sigma$ is the standard variation.
Pre-normalization is done using the statistics of the whole dataset, while post-normalization is done using the statistics of the current batch.

\subsubsection{Mixup}
A Mixup block \cite{mixup} is utilized, which mixes randomly selected past inputs with the current input in a small ratio. The past inputs are therefore used as a background sound, something which helps the network to learn representations only of the foreground acoustic event.


Since the acoustic features are log scaled, it is first converted to a linear scale, before the mixup is applied and then converted back to the log domain.
The mixing ratio that is used is sampled from a Uniform distribution $U(0, \alpha)$ where $\alpha$ is a hyper-parameter that is 0.4 in our experiments.

While in the original implementation of BYOL-A the main purpose of the mixup block is to discriminate between foreground and background, we find that it is also crucial to our approach by encouraging the model to learn more discriminative features between speakers.
This stems from the fact that the two paths of the BYOL-A training may contain past inputs sampled from different speakers.

\subsubsection{RRC}
Random Resize Crop is an image augmentation technique that is adapted for audio by applying it to the Mel Spectrogram of an audio segment. It can be conceived as an approximation of pitch shifting and time  stretching. 

The procedure consists of sampling a random crop from the log mel spectrogram. Given a number of frequency bins $F$ and of time frames $T$, the size of the crop area is randomly sampled as:
\begin{equation}\label{eq:rrc}
\begin{aligned}
   F_C &= \lfloor \min{(U(h_1, h_2), 1.0) \times F} \rfloor \\
   T_C &= \lfloor U(w_1, w_2) \times T \rfloor
\end{aligned}
\end{equation}
where $F_c$ and $T_C$ are the number of frequency and time frames respectively and $[h_1, h_2]$ and $[w_1, w_2]$ are the ranges of scaling that are being used for frequency and time. 
The default values that are used are $0.6$ and $1.5$ for both dimensions, which means that the new crop area may be outside of the boundaries of the original spectrogram. In this case zeros are used to fill the required area.
 
\subsubsection{Gaussian Noise}
A final augmentation is a Gaussian augmentation block that interpolates between training data and noise sampled from a normal distribution.
The purpose of this augmentation is similar to the mixup augmentation. The Gaussian noise is sampled from $N(0, 0.04)$ and added to the log-domain with the same log-exp trick as the mixup augmentation.

\subsection{Additional Augmentations for Robust Voice Cloning}

\subsubsection{Prosodic Augmentations}\label{pros}
While we observed that plain RRC and Mixup was sufficient for cloning, we found that better performance and robustness is possible by applying direct pitch shifting and duration scaling to the waveforms. 
The intuition for this idea is that prosodic variations should not affect speaker identity and thus, using them as augmentations should enable the self-supervised training to better focus on the speaker identity.

Both pitch shifting and duration scaling were implemented via the Praat Tolkit \cite{praat}.
The amount of shifting and scaling was treated as a hyper-parameter, but in general, it was observed that too much variation was detrimental to the speaker similarity of the cloned utterances. 
This is presumably because, while speaker identity should not be affected by small shifts in pitch, larger augmentations that take the speaker outside of their normal speaking range will  lessen similarity. The same applies to large differences in duration scaling.

It should be noted that these augmentations, unlike RRC, are being done directly in the waveform, and thus precede all the augmentations of the standard BYOL-A training. 

\subsubsection{External Noise}
Since the training of BYOL-a features is general purpose, we expect the acoustic conditions of each utterance to be present in the representations.
Since speaker identity is invariant to noises that may exist in the dataset, this is undesirable for our task.
In order to make the self-supervised features more robust to the acoustic conditions, we utilized the background noise dataset from the Chime-4 challenge \cite{noise_dataset}. 

Similarly to \cite{noise_taco}, the proposed augmentation consists of adding a randomly selected piece of noise with a randomly selected SNR to the waveform. Since the noises of these dataset are sampled from sources which may be present in real audio utterances, we expect this augmentation to better isolate the speaker identity in real world applications.

\section{Experiments and Results}
\subsection{Experimental Setup}
The audio features used for pre-training the BYOL-A model are log-scaled mel spectrograms from audios resampled to 16Khz, with 64 mel spaced frequency bins, window size of 64 ms and hop size of 10ms.  
During training, one second segments are randomly extracted from the audio, with both of the encoder networks operating on the same second. When duration augmentations were used, the beginning of each second is adjusted so that it corresponds to the same point of the waveform in the two augmented views.

The dimension of the BYOL-A embeddings used, is 512. We found that increasing the embedding size beyond that, led to the TTS decoder learning to rely on the embeddings for linguistic information as well as speaker identity, which is undesirable for cloning, as traces of the content of the reference utterance can be heard in the generated voice.

For the prosodic augmentations, a pitch shifting of -1 and 1 semitones was used and a time stretching of 0.95 and 1.05 was found to yield the best results. The external noises were added with an SNR that was randomly sampled from $[5, 10, 25]$. Both of those augmentations, as well as the Gaussian augmentations, when used, were independently done on each augmented view with a probability of $50\%$.

The TTS architecture follows \cite{lpctron} with a duration predictor and a Gaussian Upsampler to replace the attention mechanism \cite{non_attentive}.
The pre-trained embeddings are concatenated with the encoder outputs, before they are fed to the predictor and the upsampler.
The synthesized features are 22 LPC features, i.e. 20 Bark-scale cepstral coefficients, the pitch period and the pitch correlation, that are then vocoded using the LPCNet Vocoder \cite{lpcnet}.


For the training of our models, we use the VCTK multispeaker dataset \cite{VCTK} which contains 108 native English speakers, with various accents, both female and male. The total number of the training sentences seen by our model is 44k.

The evaluation of the models was done on an internal crowdsourced multispeaker dataset with 196 native English speakers. For the subjective evaluations, as well as the speaker similarity metric, eight speakers (four female and four male) were randomly selected and two utterances from each speaker were used as a target for cloning. For the MCD metric, we simply randomly selected 100 utterances from the dataset. The normalization of the features before the embedding extraction was done using the statistics of the VCTK dataset, since the actual dataset from which the utterances are drawn from should not need to be available.

As a baseline we used the same non-Attentive Tacotron architecture, but with speaker embeddings extracted using the Deep learning package Resemblyzer \cite{Wan2018GeneralizedEL} trained on the VoxCeleb2 dataset \cite{Nagrani19} which contains 6,112 speakers and 1.2 M utterances. We also trained from scratch a new Resemblyzer model, only using the VCTK dataset, in order to make the comparison with our models more meaningful. The dimension of the feature embeddings for our baseline models is 256. 

Audio samples from our experiments are available at https://innoetics.github.io/publications/ssl-cloning/index.html.

\subsection{Objective Evaluation}\label{objective}
We adapt the metric s2t-same from \cite{Stanton2021SpeakerG} which measures how similar synthesized audio from a synthesized speaker is to ground truth audio from the same speaker. While in the original context, this metric was used for speakers of the training dataset, here we use it to measure the speaker fidelity for unseen speakers.

First, we extract the speaker-level d-vectors \cite{d-vectors1} \cite{d-vectors2} using the Resemblyzer package \cite{Wan2018GeneralizedEL}. We then compute the average of those d-vectors for each speaker both from the synthesized sentences and the ground truth utterances.   

This metric is then defined as:
\begin{equation}
    \underset{j}{median} \: d(V_j^{s}, V_j^t)
\end{equation}
where the distance $d$ is the cosine distance \cite{cos} $d(u_1, u_2) = 1 - \frac{u_1}{\|u_1 \|}\frac{u_2}{\|u_2 \|}$ , $V_j^{s}$, $V_j^t$ are the averaged d-vectors over all ground truth utterances of speaker $j$ and of all synthesized sentences from speaker $j$ respectively.

\begin{table}[th]
  \caption{Objective Metrics. Lower is better for both.}
  \label{tab:obj}
  \centering
  \begin{tabular}{ l | c | c | c }
    \toprule
    \multicolumn{1}{c|}{\multirow{2}{*}{\textbf{Method}}} & 
                                         \multirow{2}{3.8em}{\centering\textbf{s2t-same clean}} & \multirow{2}{3.8em}{\centering\textbf{s2t-same noisy}} & \multirow{2}{1.5em}{\centering\textbf{MCD}}  \\
                                         \multicolumn{1}{c|}{}
                                             & & &\\
    \midrule
    d-vectors VoxCeleb2 & $0.15$  &   $0.20$  & $4.05$            \\
    d-vectors VCTK & $0.27$  &        $0.29$  &  $4.19$        \\

        \midrule
    Mixup+RRC (BYOL-A)                  &  $0.22$   &   $0.34$  &        $7.49$  \\
    Mixup+RRC+Pros                       & $0.25$  &    $0.35$  &  $7.50$             \\
    Mixup+RRC+Noise                     & $0.23$  &     $0.31$ &  $7.71$      \\
    Mixup+RRC+Pros+Noise                 & $0.23$    & $0.28$  &        $7.85$             \\
    
    \bottomrule
  \end{tabular}
\end{table}

We evaluate this similarity metric with clean and noisy utterances as target sentences. All the noisy utterances were randomly sampled from a (unseen during training) set of noises from the Chime-4 challenge, with an SNR of 5, which was the largest SNR seen during training. 

We also use Mel Cepstral Distortion \cite{MCD} as an additional metric, which measures the similarity of two aligned audio sequences. We therefore compare the similarity of sentences from an unseen speaker, with the same sentence synthesized from our model given the embedding from the model under evaluation. In order to align the sequences, Dynamic Time Warping (DTW) is used. Only clean utterances were used for this metric.

In Table \ref{tab:obj} we can see the results of the objective metrics.
The augmentations used for each experiment are shown in the Method column where Pros denotes both of the prosody augmentations defined in \ref{pros}, and Noise denotes both the Gaussian and the Chime-4 external augmentations, since early experimentation shown that they work better in tandem. 

We see that the d-vectors pre-trained in the VoxCeleb2 outperform the other models, presumably because of their better generalization ability which stems from the fact that they were trained on a much larger dataset. 
We do observe however that our models all outperform the baseline when trained in the VCTK dataset, at least for the clean utterances.

The better performance of the baseline models when presented with noisy utterances is somewhat surprising. We conjecture that since the training algorithm for these features is based on speaker verification, the encoder learned to ignore all information besides speaker identity, thereby making it robust to noises.

We also note that our augmentations improve upon the vanilla BYOL-A algorithm in terms of the s2t-same metric, although not by much. Additionally, we observe that noise augmentations, when present, lead to less degradation of quality when using noisy utterances as reference, which justifies their inclusion. 

The MCD metric deteriorates when we use the extra augmentations. A possible explanation is that the MCD metric is not affected simply by the speaker identity but on other factors as well, such as prosodic information (duration differences for example can induce a high cost on the DTW algorithm). 
Since we nudged the embeddings to become more independent of those variations, we can explain the deterioration, by the model failing to capture the rest of the speech characteristics, something which is in general desirable for voice cloning. 
The low values of the baselines for these metric are more troubling and contradict this observation however.  


\subsection{Subjective Evaluation}
We use crowdsourcing to evaluate both the speaker similarity of the synthesized utterance to the target speaker and the overall quality of the speech from a scale from 1 to 5. We have excluded all the test pages with wrong validation utterance scores, with very low (1 or 2) natural speech scores, with the exact same score in all utterances of the page, and with average page score higher than the natural speech score. This filtering process resulted in 187 listeners left to evaluate our samples. Both clean and noisy utterances were used, where the noisy utterances were derived the same way as in Section \ref{objective}.


\begin{table}[th]
  \caption{Subjective results on unseen speakers with 95\% confidence intervals. Bold results correspond to the best model for each metric.}
  \label{tab:mos}
  \centering
  \begin{tabular}{ l | c | c}
    \toprule
     \multicolumn{1}{c|}{\multirow{2}{*}{\textbf{Method}}} & 
     \multirow{2}{4.5em}{\centering\textbf{Similarity MOS}} & 
     \multirow{2}{4.5em}{\centering\textbf{Quality MOS}} \\
     \multicolumn{1}{c|}{}
     & & \\
     \midrule
     Ground Truth & $4.19 \pm 0.04$ & $4.27 \pm 0.05$\\
     \midrule
     d-vectors VoxCeleb2 & $3.26 \pm 0.16$  & $\mathbf{3.58} \pm 0.10$           \\
     d-vectors VCTK & $3.21 \pm 0.16$ & $3.47 \pm 0.14$            \\
         \midrule
      Mixup+RRC (BYOL-A)  & $3.15 \pm 0.12$ &    $3.45\pm 0.10$               \\
      Mixup+RRC+Pros    &  $\mathbf{3.30} \pm 0.11$     &        $3.49\pm 0.10$                \\
      Mixup+RRC+Noise    &  $3.17 \pm 0.12$   &         $3.47\pm 0.10$              \\
      Mixup+RRC+Pros+Noise  & $3.03 \pm 0.12$  &        $3.48\pm 0.10$                   \\
     \bottomrule
  \end{tabular}
\end{table}

The results for regular utterances can be seen in Table \ref{tab:mos} and for noisy utterances in Table \ref{tab:mos_noisy}. We can see that both the quality and the speaker similarity of our models is comparable to the  d-vectors model trained in the VoxCeleb2 dataset and usually outperforms or is very close to the d-vectors trained on the VCTK dataset. We also outperform both baselines when using the pre-trained model with the prosody augmentations with clear utterances.

The inclusion of noise augmentations is further justified, as they are the best performing of our models when evaluated on the noisy utterances. They also perform quite similarly to the baseline models.

\begin{table}[th]
  \caption{Subjective results on unseen speakers with noisy utterances with 95\% confidence intervals. Bold results correspond to the best model for each metric}
  \label{tab:mos_noisy}
  \centering
  \begin{tabular}{ l | c | c}
     \toprule
     \multicolumn{1}{c|}{\multirow{2}{*}{\textbf{Method}}} & 
     \multirow{2}{4.5em}{\centering\textbf{Similarity MOS}} & 
     \multirow{2}{4.5em}{\centering\textbf{Quality MOS}} \\
     \multicolumn{1}{c|}{} & & \\
     \midrule
     Ground Truth & $4.19 \pm 0.04 $ & $4.27 \pm 0.05$\\
     \midrule
     d-vectors VoxCeleb2 & $3.03 \pm 0.16$ & $\mathbf{3.52}\pm 0.10$           \\
     d-vectors VCTK & $\mathbf{3.14 }\pm 0.17$ & $3.38\pm 0.15$            \\
     \midrule
      Mixup+RRC (BYOL-A)  & $2.98 \pm 0.13$ & $3.38\pm 0.10$            \\
      Mixup+RRC+Pros      & $2.97 \pm 0.13$  &      $3.37\pm 0.11$              \\
      Mixup+RRC+Noise     &  $3.04 \pm 0.13$  &         $3.47\pm 0.10$        \\
      Mixup+RRC+Pros+Noise  & $3.12 \pm 0.13$  &    $3.38\pm 0.10$                    \\
     \bottomrule
  \end{tabular}
\end{table}
\section{Conclusions}
We present a new voice cloning architecture, based on self-supervised features that are pre-trained on an unlabeled dataset. We show that it is close in performance to the baseline of speaker features that are pre-trained on speaker verification tasks even when used on a fraction of the dataset, and even without having any information about the speaker identity of the training sentences. We also further extend the set of augmentations applied in the self-supervised algorithm that both improve the cloning performance and quality and make the model more robust to noise in the target utterances. Future work on this topic could include further exploration of augmentations to improve performance, or the utilization of self-supervised features for different TTS tasks such as prosody transfer or emotional speech synthesis.

\clearpage

\bibliographystyle{IEEEtran}

\bibliography{mybib}

\end{document}